\documentclass[final,5p,times,twocolumn]{elsarticle}

\usepackage{amssymb}
\journal{Physics Letters A}

\begin{document}

\begin{frontmatter}



\title{A mixed SOC-turbulence model for nonlocal transport and L\'evy-fractional Fokker-Planck equation}


\author{Alexander~V.~Milovanov}
\address{ENEA National Laboratory, Centro~Ricerche~Frascati, I-00044 Frascati, Rome, Italy\\
Department of Space Plasma Physics, Space Research Institute, Russian Academy of Sciences, 117997 Moscow, Russia}

\author{Jens~Juul~Rasmussen}
\address{Physics Department, Technical University of Denmark, DK-2800 Kgs. Lyngby, Denmark}

\begin{abstract}
The phenomena of nonlocal transport in magnetically confined plasma are theoretically analyzed. A hybrid model is proposed, which brings together the notion of inverse energy cascade, typical of drift-wave- and two-dimensional fluid turbulence, and the ideas of avalanching behavior, associable with self-organized critical (SOC) behavior. Using statistical arguments, it is shown that an amplification mechanism is needed to introduce nonlocality into dynamics. We obtain a consistent derivation of nonlocal Fokker-Planck equation with space-fractional derivatives from a stochastic Markov process with the transition probabilities defined in reciprocal space. The hybrid model observes the Sparre Andersen universality and defines a new universality class of SOC.
\end{abstract}


\end{frontmatter}

% \linenumbers

\section{Introduction} In magnetically confined plasmas, perturbative experiments \cite{Mantica} with plasma edge cooling and heating power modulation reveal anomalously fast transport of edge cold pulses to plasma core, not compatible with major diffusive time scales \cite{Pulse}. The rapid edge cooling often leads to an increase in temperature in the plasma core by reversal of the sign of the perturbation \cite{Mantica_etal}. Cold pulse reversal is also related to a rather abnormal effect of spontaneous rotation reversal, where the core rotation velocity of plasma changes sign spontaneously and in the absence of local sources \cite{Rice}. It was argued that diffusive transport models based on Fick's second law and local density gradients{\footnote {The Fick paradigm states that the internal fluxes are described by a set of local transport coefficients $-$ diffusivities or conductivities $-$ related to the local thermodynamic forces which induce the fluxes through Fick's law. Models based on these assumptions are referred to as local models.}} were problematic to accommodate the observed behaviors and that there was a connection \cite{Pulse,Castillo2,Pulse+} between the cold pulse problem and the phenomena of nonlocal transport described by kinetic equations with fractional derivatives in space \cite{Nature,Klafter,Report}. The latter are integro-differential operators \cite{Klafter,Samko} incorporating in a mathematically appealing fashion the key signatures of non-Gaussianity and long-range dependence beyond the restrictive assumptions of locality and lack of correlations underlying the standard diffusive style paradigm. 

A criticism raised against those models, however, is that their application in the realm of fusion research has been based on phenomenological arguments and heuristic assumptions rather than microsopic equations of the motion of charged particles. More so, it is not clear which plasma conditions, values of parameters, and key physics ingredients behind the plasma confinement one really needs in order to generate nonlocal behavior. For instance, shall the familiar drift-wave turbulence of the Hasegawa-Wakatani (HW) type \cite{HW} accommodate nonlocal transport conformally with the fractional diffusion models? Another important issue is validation of nonlocal equations from microscopic dynamics of diffusing charged particles. It is our aim in the present work to obtain progress over these topics. 

In what follows, we address the nonlocal transport problem from a more fundamental perspective, namely, by advancing the concept of nonlocal transport driven by a stochastic noise process of the L\'evy type. The key idea behind this approach is inspired by the early work of Chechkin and Gonchar \cite{Gonchar}, but with a different derivation using the notion of transition probability in reciprocal space. We show that nonlocal behavior does not really occur in the typical HW setup, if only at the margins of validity by stretching values of parameters into the regimes with strong nonlinearity. Then a consistent picture of the transport is found in the realm of ``complexity" coupling \cite{AJ} between the phenomena of drift-wave turbulence and self-organized criticality (SOC) dynamics \cite{Bak}. 

It should be noted that there is no commonly agreed upon the use of the term {\it complexity}. By this we shall mean back-reaction between two or more dynamical degrees of freedom in essentially a nonlinear context. Topical examples of complex behavior in fusion plasma include coupling between the density and the potential fluctuations in the HW picture of drift-wave turbulence \cite{HW}; the coupled drift-wave and zonal-flow turbulent system \cite{DM2005}; blob generation from drift waves and interchange instabilities \cite{Ippolito}; and recently observed in the TEXTOR experiments the internal-kink (``fishbone") and edge-localized mode (ELM) coupling \cite{Meijere}. A unifying feature among these phenomena is that they reveal the presence of strong nonlinearity beyond perturbation theory approaches. No wonder that the dynamics of complex systems has been considered difficult to investigate \cite{UFN,Nicolis,Milo13,Sibani}. 

The theoretical concept of SOC has been initially applied by Bak {\it et al.} \cite{Bak} to describe sandpile avalanches at a critical angle of repose, and has been generalized to nonlinear dissipative systems that are driven in a ``critical" state. The phrase ``self-organized" implies that the system reaches the critical state without any tuning of parameters. It has been slowly accepted that SOC occurs through a nonlinear feedback mechanism triggering intermittent, avalance-like transitions between different metastable states \cite{Kadanoff,Sor}. Before that acceptance, the notion of SOC has been widely discussed and debated in the literature \cite{MF98}. For a system in a SOC state local avalanche analysis reveals that the probability distributions of avalanche size and duration are approximately scale-free, but deviations from perfect scaling and their relation to the finite system size have size scaling and multifractal scaling. These topics are summarized in a recent book \cite{Ash2013} and in a review \cite{SSRv}. 

Much theoretical and numerical effort has been invested to discriminate between the theoretical concepts of turbulence and SOC and to identify a set of observable properties which are the unique fingerprint of SOC \cite{Freeman,Watkins,PRB07,Ash2012}. In magnetically confined fusion plasma, SOC has been proposed as an {alternative} to the turbulence theoretical framework to explain and control the anomalous particle and heat transport across the magnetic field lines and the extreme transport events that may have destructive effect on very expensive plasma-facing components \cite{Newman96,Politzer,Tokunaga}. Here by contrast with this way of thinking we suggest that the phenomena of turbulence and SOC are {\it not} really separable in tokamaks in the regime of strong nonlinearity and that there is a theoretical possibility that the turbulence fuels the avalanching dynamics due to SOC through inverse cascade of the energy, giving rise to transport events of anomalously large size beyond the range of predictability of the ``conventional" SOC. We envisage this fueling process as amplification of the SOC avalanches by the turbulence \cite{AJ}. %It is at this point, where the phenomena of complexity come into play.

The purpose of this paper is to address this new form of complexity phenomenon, the SOC-turbulence coupling, which explicitly takes into account the back-reaction of the inverse energy cascade on SOC. We expect these coupling phenomena to universally occur in two-dimensional fluid (as well as fluid-like, such as the drift-wave) turbulence in the presence of a nonlinear feedback mechanism generating SOC. Then the processes of amplification taking place will manifest themselves in the form of algebraic tails on top of the typical log-normal behavior of the probability distribution function of the flux-surface averaged transport. ``Algebraic" means that these tails pertain to a category of processes described by the statistics of the L\'evy type, thus paving the way to the derivation of fractional transport models by standard methods. Physically, the algebraic tails shall represent outstanding transport events, which we associate with large intermittent bursts of transport. Indeed big events falling off the usual transport metrics in magnetically confined plasma have been reported in tokamak phenomenology \cite{Xu10}. The typical examples of this behavior include ELMs, as they are now commonly known \cite{ELM}, and blob-filaments, which are magnetic-field-aligned plasma structures that are considerably denser than the surrounding background plasma and are highly localized in the directions perpendicular to the equilibrium magnetic field lines. In experiments and simulations, intermittent filaments are often formed near the boundary between open and closed field lines, and seem to arise in theory from saturation process for the dominant edge instabilities and turbulence. Blob transport is of interest from a fundamental scientific perspective, since it is a general phenomenon occurring in nearly all plasmas \cite{Ippolito}.

The paper is organized as follows. We shall first derive a L\'evy-fractional Fokker-Planck equation from a generic Markov stochastic process in configuration space. Then we shall discuss the basic physics implications of this derivation in brief. Finally, we set the model in a more general context and address the connections with turbulence- and SOC-associated phenomena. We argue that the phenomena of SOC-turbulence coupling observe the Sparre Andersen universality \cite{SA53} and we employ the concept of the Galton-Watson chain process to theoretically predict the exponent of fractional differintegration over the space variable. Our results conform well with the statistics of tokamak plasma fluctuations.   

\section{L\'evy-Fractional Fokker-Planck equation} We work with a Markov (memoryless) stochastic process defined by the evolution equation\footnote{The case of velocity-space transport, though conceptually similar, is not discussed here. Nor do we discuss processes with trapping, leading to slow diffusion and fractional time derivatives in the end.}
\begin{equation}
f(x, t+\Delta t) = \int_{-\infty}^{+\infty} f(x-\Delta x, t) \psi (x, \Delta x, \Delta t)d\Delta x,
\label{1} % Eq.~(\ref{1})
\end{equation}
where $f (x, t)$ is the probability density of finding a particle (random walker) at time $t$ at point $x$ and $\psi (x, \Delta x, \Delta t)$ is the transition probability density of the process. Note that the ``density" $\psi (x, \Delta x, \Delta t)$ is defined with respect to the increment space characterized by the variable $\Delta x$. It may include a parametric dependence on $x$, when non-homogeneous systems are considered. Here, for the sake of simplicity, we restrict ourselves to the homogeneous case, and we omit the $x$ dependence in $\psi (x, \Delta x, \Delta t)$ to obtain
\begin{equation}
f(x, t+\Delta t) = \int_{-\infty}^{+\infty} f(x-\Delta x, t) \psi (\Delta x, \Delta t)d\Delta x.
\label{2} % Eq.~(\ref{2})
\end{equation} 
Then $\psi (\Delta x, \Delta t)$ defines the probability density of changing the spatial coordinate $x$ by a value $\Delta x$ within a time interval $\Delta t$, independently of the running $x$ value. The integral on the right of Eq.~(\ref{2}) is of the convolution type. In the Fourier space this becomes
\begin{equation}
\hat f(k, t+\Delta t) = \hat f(k, t) \hat \psi (k, \Delta t),
\label{3} % Eq.~(\ref{3})
\end{equation} 
where the integral representation  
\begin{equation}
\hat \psi (k, \Delta t) = \hat \mathcal{F} \{\psi (\Delta x, \Delta t)\} \equiv \int_{-\infty}^{+\infty} \psi (\Delta x, \Delta t) e^{ik\Delta x} d\Delta x
\label{Fourier} % Eq.~(\ref{Fourier})
\end{equation} 
has been used for $\hat \psi (k, \Delta t)$, and similarly for $\hat f(k, t)$. Letting here $k\rightarrow 0$, it is found that 
\begin{equation}
\lim_{k\rightarrow 0}\hat \psi (k, \Delta t) = \int_{-\infty}^{+\infty} \psi (\Delta x, \Delta t) d\Delta x.
\label{F2} % Eq.~(\ref{F2})
\end{equation} 
The improper integral on the right hand side is nothing else than the probability for the space variable $x$ to acquire {\it any} increment $\Delta x$ during time $\Delta t$. For memoryless stochastic processes without trapping, this probability is immediately seen to be equal to 1 (the diffusing particle takes a displacement anyway in any direction on the $x$-axis), given that the time interval $\Delta t$ is longer than the characteristic width of the driving-force spikes. Thus,
\begin{equation}
\lim_{k\rightarrow 0}\hat \psi (k, \Delta t) = 1.
\label{F2+} % Eq.~(\ref{F2+})
\end{equation} 
We consider $\hat \psi (k, \Delta t)$ as the average time-scale- and wave-vector-dependent transition ``probability" or the characteristic function of the stochastic process in Eq.~(\ref{2}). In general, $\hat \psi (k, \Delta t)$ can be due to many independent, co-existing processes, each characterized by its own, ``partial" transition probability, $\psi_j (k, \Delta t)$, $j=1,\dots n$, making it possible to expand
\begin{equation}
\hat \psi (k, \Delta t) = \prod_{j=1}^n \hat \psi_j (k, \Delta t).
\label{Prod} % Eq.~(\ref{Prod})
\end{equation} 
We should stress that, by their definition as Fourier integrals, $\hat \psi_j (k, \Delta t)$ are given by complex functions of the wave vector $k$, and that their appreciation as ``probabilities" has the only purpose of factorizing in Eq.~(\ref{Prod}). Even so, with the aid of Eq.~(\ref{F2}) above, this factorized form is justified via the asymptotic matching procedure in the limit $k\rightarrow 0$. In practice, aiming at the prospective fluid and plasma applications, it is sufficient to address a simplified version of Eq.~(\ref{Prod}), where just two co-existing key processes are included $-$ one corresponding to a white noise-like process, which we shall mark by the index $L$; and the other one, corresponding to a regular convection process, such as a zonal flow or similar, which we shall mark by the index $R$. We have, accordingly,  
\begin{equation}
\hat \psi (k, \Delta t) = \hat \psi_L (k, \Delta t) \hat \psi_R (k, \Delta t).
\label{Prod2} % Eq.~(\ref{Prod2})
\end{equation}
These settings correspond to a set of Langevin equations
\begin{equation}
dx/dt = v;~dv/dt = -\eta v + F_R + F_L (t),
\label{Lvin} % Eq.~(\ref{Lvin})
\end{equation}
where $\eta$ is the fluid viscosity; $F_R$ is the regular force; and $F_L (t)$ is the fluctuating (noise-like) force. We take $F_L (t)$ to be a white L\'evy noise with L\'evy index $\mu$ ($1 < \mu\leq 2$). By white L\'evy noise $F_L (t)$ we mean a stationary random process, such that the corresponding motion process, i.e., the time integral of the noise, $L (\Delta t) = \int_t^{t+\Delta t} F_L (t^{\prime}) dt^{\prime}$, is a symmetric $\mu$-stable L\'evy process with stationary independent increments and the characteristic function  
\begin{equation}
\hat \psi_L (k, \Delta t) = \exp (-D_\mu |k|^\mu \Delta t) \sim 1 - D_\mu |k|^\mu \Delta t.
\label{GCLT} % Eq.~(\ref{GCLT})
\end{equation}
The last term gives an asymptotic inverse-power distribution of jump lengths 
\begin{equation}
\chi (\Delta x) \sim |\Delta x|^{-1-\mu}.
\label{Jump-l} % Eq.~(\ref{Jump-l})
\end{equation}
The constant $D_\mu$ constitutes the intensity of the noise. As is well-known, the characteristic function in Eq.~(\ref{GCLT}) generates L\'evy flights \cite{Klafter,Report}. 

Focusing on the regular component of the force field, $F_R$, it is convenient to choose the corresponding transition probability in the form of a plane wave, i.e.,
\begin{equation}
\hat \psi_R (k, \Delta t) = \exp (iuk\Delta t) \sim 1 + iuk\Delta t.
\label{Plane} % Eq.~(\ref{Plane})
\end{equation}
Here, $u$ is the speed of the ``wave," which is decided by convection. One evaluates this speed by neglecting the term $dv/dt$ in Langevin equations~(\ref{Lvin}) to yield $u=F_R/\eta$. It is noted that the general condition in Eq.~(\ref{F2+}) is clearly satisfied for both the L\'evy processes and the stationary convection, emphasizing the Markov property and the absence of trapping. Putting all the various pieces together, one readily obtains 
\begin{equation}
\hat \psi (k, \Delta t) = \exp (-D_\mu |k|^\mu \Delta t + ik F_R \Delta t / \eta).
\label{Tog} % Eq.~(\ref{Tog})
\end{equation} 
The next step is to substitute this into Eq.~(\ref{3}), and to allow $\Delta t \rightarrow 0$. Then, Taylor expanding on the left- and right-hand sides in powers of $\Delta t$, and keeping first non-vanishing orders, in the long-wavelength limit $k\rightarrow 0$ it is found that 
\begin{equation}
\frac{\partial}{\partial t} \hat f (k, t) = \left[-D_\mu |k|^\mu + ik F_R / \eta \right] \hat f (k, t).
\label{9} % Eq.~(\ref{9})
\end{equation} 
When inverted to configuration space, the latter equation becomes
\begin{equation}
\frac{\partial}{\partial t} f (x, t) =  \left[D_\mu \frac{\partial^\mu}{\partial |x|^\mu} - \frac{1}{\eta} \frac{\partial}{\partial x} F_R \right] f (x, t),
\label{Inv} % Eq.~(\ref{Inv})
\end{equation} 
where the symbol $\partial^{\mu} / \partial |x|^\mu$ is defined by its Fourier transform as  
\begin{equation}
\hat \mathcal{F} \Big\{\frac{\partial^\mu}{\partial |x|^\mu}f (x, t)\Big\} = -|k|^\mu \hat f (k,t),
\label{Def} % Eq.~(\ref{Def})
\end{equation} 
and we have followed the usual convention \cite{Klafter} of suppressing the imaginary unit in Fourier space. In the foundations of fractional calculus \cite{Samko} it is shown that, for $1 <\mu < 2$, 
\begin{equation}
\frac{\partial^\mu}{\partial |x|^\mu} f (x, t) = \frac{1}{\Gamma_\mu}\frac{\partial^2}{\partial x^2} \int_{-\infty}^{+\infty}\frac{f (x^\prime, t)}{|x-x^\prime|^{\mu - 1}} dx^\prime,
\label{Def+} % Eq.~(\ref{Def+})
\end{equation} 
where $\Gamma_\mu = - 2\cos(\pi\mu/2)\Gamma(2-\mu)$ is a numerical normalization parameter. One sees that $\partial^\mu / \partial |x|^\mu$ is an integro-differential operator, which has the analytical structure of ordinary space differentiation acting on a Fourier convolution of the function $f (x,t)$ with a power-law. It interpolates between a pure derivative and a pure integral, and is often referred to as the fractional Riesz operator. By its definition, the Riesz operator can conveniently be considered as a normalized sum of left and right Riemann-Liouville derivatives on the infinite axis. It is this operator, which incorporates the nonlocal properties of the transport. In the Gaussian limit $\mu = 2$, the Riesz operator reduces to the conventional Laplacian, so that local behavior is recovered. Relating $F_R$ to an external potential field, $F_R = -V^{\prime}(x)$, we are led to the following fractional Fokker-Planck equation, or FFPE (e.g., Refs. \cite{Gonchar,Chechkin}; reviewed in Refs. \cite{Klafter,Klafter2004}) 
\begin{equation}
\frac{\partial}{\partial t} f (x, t) =  \left[D_\mu \frac{\partial^\mu}{\partial |x|^\mu} + \frac{1}{\eta} \frac{\partial}{\partial x} V^{\prime}(x)\right] f (x, t).
\label{FFPE} % Eq.~(\ref{FFPE})
\end{equation} 
In writing FFPE with the spatial dependence in $V^{\prime}(x)$ we have also assumed that the scales of the $F_R$ variation are smooth compared with the fluctuation noise-like scales. 

FFPE~(\ref{FFPE}) can alternatively be derived using as a starting point the set of Langevin equations~(\ref{Lvin}) instead of the evolution equation in Eq.~(\ref{2}). The advantage of Langevin approach lies in the straightforward way of including the force terms due to the various driving processes in the medium. In this connection, we also note that the study of nonlocal transport in terms of Langevin equations with a L\'evy noise and the corresponding generalized Fokker-Planck equation containing fractional derivatives in space has been also suggested by Fogedby \cite{Fog} and Jespersen {\it et al.} \cite{Jesper}. 

We should stress that the introduction of an $x$-dependent force field $F_R (x) = -V^{\prime}(x)$ in place of the constant force field in Eq.~(\ref{Inv}) destroys the spatial homogeneity of the transfer statistics implied by the transfer kernel in Eq.~(\ref{2}). Even so, this extension to non-homogeneous systems with the spatial asymmetry due to the force $F_R (x)$ could be employed under the condition that the terms determining the jump length $|x-x^\prime|$ separate from the coordinate dependence in $F_R (x)$, implying that the force is calculated at the arrival site $x$ and not at the departure site $x^\prime$. Technically, the separation of terms can be implemented based on the generic functional form \cite{Barkai} of the memory kernel, using the Heaviside step function to ascribe the dependence on the jump length. More so, implementing a similar convention regarding the arrival site, the assumption that the intensity of the L\'evy noise $D_\mu$ does not depend on $x$ can be relaxed. At this point, one confronts non-homogeneous transport models with $D_\mu = D_\mu (x)$. The non-homogeneity of the noise term can be an inherent property of fluctuations driving the transport and can occur naturally as a consequence of nonlinear interaction between the components of the force field \cite{EPL_Iom}. Indeed it is found in the mean-field approximation that the nonlinear interactions in spatially extended systems can result in a range-dependent transport and that the behavior does not possess a characteristic scale \cite{EPL_Iom,Iomin}. Also in tokamak geometry, the range-dependence of the driving noise term can be motivated by the phenomena of asymmetric radial spreading of fluctuations; implying that the turbulence itself is a transported quantity and can penetrate into linearly stable regions of the plasma \cite{Pulse+}. Then a meaningful toy-model with competition between nonlocality and non-homogeneity is represented by  
\begin{equation}
\frac{\partial}{\partial t} f (x, t) =  \left[K_\mu \frac{\partial^\mu}{\partial |x|^\mu} |x|^{-\theta} + \frac{1}{\eta} \frac{\partial}{\partial x} V^{\prime}(x)\right] f (x, t),
\label{FFPE_x} % Eq.~(\ref{FFPE_x})
\end{equation} 
where $K_\mu$ is a constant and does not depend on $x$, and $\theta$ ($\theta > 0$) absorbs in a simple scaling the parameters of the interaction. One sees that the range-dependence appears in an anomalous dispersion law $t\sim |x|^{\mu + \theta}$; where the behavior is superdiffusive for $\theta < 2 - \mu$; and subdiffusive otherwise. Thus, the range dependence coexisting with nonlocal derivatives slows down the transport within some limits, with a room for subdiffusive scaling in the parameter range $\theta > 2 - \mu$. This result challenges the conventional picture of L\'evy flights as a paradigmatic model for superdiffusion. We note in passing that the fractional derivatives due to the fluctuating noise-like force appear in the generalized diffusion term, but not really in the convection term involving the potential force field. This property stemming from the factorization in Eq.~(\ref{Prod}) using $k\rightarrow 0$ can be also demonstrated based on continuous time random walk schemes \cite{Barkai}. Mathematically, nonlocal equations with the range-dependent $D_\mu (x) \propto |x|^{-\theta}$ have been considered by Srokowski \cite{Srok}, where one can also find their solutions in terms of the Fox $\mathrm{H}$-function. In the remainder of this paper we shall assume a uniform on the large scales fluctuation background and we omit correspondingly the scaling dependence in $D_\mu$ consistently with the assumptions of homogeneity behind Eq.~(\ref{2}) above.     

Equation~(\ref{FFPE}) can be extended, so that it includes local transport due to e.g., collisions, in addition to the nonlocal transport processes discussed above. The key step is to observe that collisions will generate a white noise of the Brownian type, whose characteristic function is just a Gaussian, and is obtained from the general L\'evy form~(\ref{GCLT}) for $\mu\rightarrow 2$. We note in passing that the Gaussian law, too, belongs to the class of stable distributions, but it will be the only one to produce finite moments at all orders. When the L\'evy and Brownian noises are included as independent elements to the dynamics, the transition probability in Eq.~(\ref{Prod}) will again factorize, and will acquire, in addition, a Gaussian factor $\hat \psi_G (k, \Delta t) = \exp (-D k^2 \Delta t)$, where $D$ has the sense of collisional diffusion coefficient. Then Eq.~(\ref{Tog}) will generalize to   
\begin{equation}
\hat \psi (k, \Delta t) = \exp (-D_\mu |k|^\mu \Delta t -D k^2 \Delta t + ik F_R \Delta t / \eta),
\label{CLT+} % Eq.~(\ref{CLT+})
\end{equation}  
from which a FFPE incorporating both the fractional Riesz and the usual Laplacian operators 
\begin{equation}
\frac{\partial}{\partial t} f (x, t) =  \left[D_\mu \frac{\partial^\mu}{\partial |x|^\mu} + D \frac{\partial^2}{\partial x^2} + \frac{1}{\eta} \frac{\partial}{\partial x} V^{\prime}(x) \right] f (x, t)
\label{FFPE+} % Eq.~(\ref{FFPE+})
\end{equation} 
can be deduced for small $k$. In the applications \cite{VanM2004,Carr2006}, it is convenient to think of the L\'evy noise $F_L (t)$ as involving a critical threshold condition in that the intensity $D_\mu$ is only non-zero off a certain critical value of the average gradient generating the instabilities, and vanishes otherwise. Then the FFPE in Eq.~(\ref{FFPE+}) readily switches between local (e.g., collisional, as well as Gaussian quasi-linear) transport in the parameter range of sub-critical behavior, and nonlocal (L\'evy style) transport above the criticality. FFPEs of the type~(\ref{FFPE+}) with the combination of L\'evy and ordinary diffusion have been also discussed in connection with the dynamics of protein fast-folding and the motion of excitations and proteins along polymer chains \cite{Ralf1,Ralf2}. 

Our next point concerns the issue of boundary conditions for nonlocal kinetic equations discussed above. Of interest here is an {\it absorbing} boundary, which we associate with the edge of a magnetically confined plasma. We argue below that it is the absorbing boundary, which accommodates the phenomena of SOC. So if we place an absorbing boundary at $x=0$, with an initial condition defined along the positive semi-axis $x > 0$, then the solution necessarily has to vanish on the negative semi-axis. Then the nonlocal integration from $-\infty$ to $+\infty$ in the corresponding Riesz operator should be reduced to the positive semi-axis, yielding
\begin{equation}
\left[\frac{\partial}{\partial t} - \frac{1}{\eta} \frac{\partial}{\partial x} V^{\prime}(x) \right] f (x, t) = \frac{\partial^2}{\partial x^2} \Psi (x, t) + D \frac{\partial^2}{\partial x^2} f (x, t), 
\label{Sparre} % Eq.~(\ref{Sparre})
\end{equation} 
where 
\begin{equation}
\Psi (x, t) = \frac{D_\mu}{\Gamma_\mu} \int_{0}^{+\infty}\frac{f (x^\prime, t)}{|x-x^\prime|^{\mu - 1}} dx^\prime, 
\label{Sparre+} % Eq.~(\ref{Sparre+})
\end{equation} 
and we have moved the convection term to the left-hand-side for convenience. Following Chechkin {\it et al.} \cite{Chechkin2003}, one finds in the presence of an absorbing boundary that FFPE~(\ref{Sparre}) with improper integration in Eq.~(\ref{Sparre+}) correctly phrases the first passage time density problem \cite{Klafter2004} for L\'evy flights. It will moreover observe the Sparre Andersen universality  \cite{SA53} that the first passage time density decays as $\sim t^{-3/2}$ after $t$ time steps ($t\rightarrow +\infty$). We consider this universality as a characteristic property of the nonlocal transport model.    

\section{To be, or not to be \cite{Shak}} We have seen in the above that the statistical case of nonlocal transport stems from a driving noise-process of the L\'evy type, whereas regular convection acts as to introduce an external potential field to L\'evy flights, contained in the $V(x)$ dependence. On the one hand, this spotlights the basic physics implications of the fractional derivative operator occurring in FFPE~(\ref{FFPE}). On the other hand, it casts doubts on its relevance to the classical picture of drift-wave- and two-dimensional fluid turbulence, as well as to the classical picture of SOC. The main elements of concern consist in the following. In SOC, one is interested in how long-time correlated dynamics will develop via local couplings between the many degrees of freedom leading to complex patterns \cite{Bak,Ash2013}. Then the assumed next-neighbor character of lattice interactions in the vicinity of criticality will be incompatible with nonlocal space differentiation, so that the correlations that are long-ranged enter through nonlocal differentiation over the time, rather than the space, variable, thus preserving the local structure of the Laplacian \cite{Milo13,PRE09,NJP}. 

Further concerns come from non-observation, over a statistically significant range, of algebraic tails in direct numerical simulations of electrostatic drift-wave, as well as electromagnetic drift-Alfv\'en, turbulence; also supported by the available experimental evidence. A particle going with a strongly turbulent flow will experience a sequence of flights, and we can reasonably expect that the distribution of flight lengths will approximately follow the average particle flux distribution. The latter is obtained as a flux-surface integral of the convected density, i.e., $\Gamma = \int d\sigma \left[\tilde n u_{E\times B}\right]/\int d\sigma$, where $\tilde n$ and $u_{E\times B}$ are the density fluctuations and the $E\times B$ velocity, respectively. Then a statistics of the L\'evy type will imply that the probability density function of the averaged transport $\Gamma$ will exhibit an algebraic tail, i.e., $\chi (\Gamma) \sim \Gamma^{-1-\mu}$, whereas the simulations have revealed an exponential tail \cite{Basu,PLA,Garcia,Garcia+}.     

Based on this reasoning, we are led to infer that the fractional FFPE in Eq.~(\ref{FFPE}) is neither consistent with the classical two-dimensional fluid turbulence approach including its drift-wave and drift-Alfv\'en counterparts, nor classical SOC approach built on the assumptions of locality and next-neighbor interactions. Thus, a more intricate gateway for nonlocal transport should be agreed. Here, we suggest that nonlocality comes into play as a result of amplification (and amplification of amplification) of the SOC avalanches in the presence of inverse energy cascade in a two-dimensional turbulent flow. That means that in fact {\it two} fundamental ingredients, operating in concert, and essentially on an equal footing, are needed to generate nonlocal transport: fluid-like behavior with the inverse turbulent cascade on the one hand, and avalanching dynamics involving SOC on the other hand. This idea leads directly to the L\'evy statistics, as we now proceed to show.    

\section{Turbulent amplification process} In the combined SOC-turbulence scenario, which we consider, a guiding role is attributed to the usual picture of eddies and eddy-induced transport associated with plasma instabilities. A paradigmatic framework to understand and explain the phenomenon from first principles is drift-wave turbulence modeled by the well-known Hasegawa-Wakatani (HW) equations, as direct numerical investigations show \cite{Basu,PLA,JJR}. Nonlinearly, in a magnetic confinement geometry, the eddy-induced transport reduces the slope of the average profile where the vorticity is maximal, and, at the same time, steepens it in its nearby vicinity, thus increasing the instability in the next radial location. This displacement of the instability is an avalanche in that the step in the gradient moves radially outward, creating an unstable propagating front.\footnote{A report on observation and quantitative characterization of avalanche events in a magnetically confined plasma can be found in Ref. \cite{Politzer}. It was argued that the evidence that avalanche events are present in the plasma is ``strong," and that the observations are qualitatively similar to results of modeling calculations based on drift-wave turbulence.} However, because drift-wave turbulence is essentially two-dimensional, there exists an inverse cascade of the energy which is associated with the phenomena of eddy merging and the formation of large-scale coherent structures in the strongly turbulent flow. To this end, the propagation of the unstable front becomes a combined effect due to the next-radial generation of the off-spring eddies and their merging with the ever-growing mother eddy. One sees that turbulence will act as to amplify the avalanches by fueling them with more free energy via the inverse cascade. The process will stop when excess energy and particles are eventually let out through boundaries, thus reducing the slope of the average profile on the system-size scales. 

We employ the HW model \cite{HW,JJR} for two-dimensional electrostatic drift-wave turbulence driven by the resistive instability 
\begin{equation}
\frac{\partial}{\partial t} \tilde n + \frac{\partial}{\partial y}\varphi + \{\varphi, \tilde n\} = - \frac{1}{\delta} (\tilde n - \varphi),
\label{HW-a} % Eq.~(\ref{HW-a})
\end{equation} 
\begin{equation}
\frac{\partial}{\partial t} \nabla^2\varphi + \{\varphi, \nabla^2\varphi\} = - \frac{1}{\delta} (\tilde n - \varphi) + \kappa \nabla^4 \varphi.
\label{HW-b} % Eq.~(\ref{HW-b})
\end{equation} 
Here the Poisson bracket $\{\varphi, g\} = \hat {\bf z} \cdot \left[\nabla\varphi \times \nabla g\right]$ is used to denote the nonlinear terms originating from advection with the $E\times B$ drift; $E = -\nabla\varphi$ is the intensity of electrostatic field; $\varphi$ is the electrostatic potential; $\nabla^2\varphi$ is the vorticity, that is, the curl of $E\times B$; and we have replaced the parallel derivatives with an effective parallel wavelength. A damping term $\kappa \nabla^4 \varphi$ is added to Eq.~(\ref{HW-b}) to dissipate vorticity on the small scales. It is assumed that only one mode parallel to the homogeneous outer magnetic field ${\bf B} = B_0 \hat {\bf z}$ is excited, so that all differential operators work in the $(x,y)$-plane. We may identify the $x$-coordinate with the radial direction and the $y$-coordinate with the poloidal direction, as a portion of the plasma-edge in a toroidal geometry is represented. Further $\delta = 1/k_{||}^2L_{||}^2$ is adiabaticity parameter and characterizes the deviation between the potential and the density fluctuations in the HW model. In a basic theory of drift waves it is shown that this deviation leads to an instability with a maximum linear growth rate $\gamma_{\rm L} \approx \delta / 8$ at $k=\nabla^{-1}\approx 1$ \cite{JJR}. It is therefore convenient to think of $\delta$ as of driving rate for the turbulence. The inverse adiabaticity parameter, $1/\delta$, characterizes the coupling strength (in the complexity sense of the wording) between the two interacting degrees of freedom represented by respectively the $\tilde n$ and $\varphi$ dependencies. It should be noted that $\delta$ absorbs via the scale length $L_{||} = (L_n T_e / m_e c_s \nu_{ei})^{1/2}$ the parameters of parallel dynamics; it also contains through the electron-ion collisional frequency $\nu_{ei}$ the parallel resistivity. Here, $L_n$ is the scale length of the perpendicular background density gradient, i.e., $L_n^{-1} = \nabla \tilde n / n_0$. The normalization scales are $c_s / \Omega_i$ for lengths perpendicular to the magnetic field and $L_n / c_s$ for the times, where $\Omega_i$ is the ion cyclotron frequency and $c_s = \sqrt{T_e / m_i}$ is the sound speed. Although the mathematical structure of the HW equations~(\ref{HW-a}) and~(\ref{HW-b}) is clear and simple, their solutions for turbulent flows are immensely complex and can only be obtained approximately through numerical computation. This absence of simple solutions is due to the nonlinear nature of the equations, giving rise to mode coupling, build-up of correlations and the formation of coherent vortical structures in the strongly turbulent flow \cite{PLA,JJR}. 

We are now in position to obtain a simple criterion for the onset of avalanching dynamics and its amplification by inverse cascade of the energy. This criterion shall use the idea of separation between wave-like motions and nonlinear structures. In drift-wave turbulence, the so-called Rhines length, $\lambda_{\rm Rh}$, determines the upper bound on the size of vortical structures in the flow. The Rhines length, originally introduced in geophysical fluid turbulence \cite{Rhines}, and later applied to drift-wave turbulence \cite{Nau}, designates the spatial scale separating vortex motion from drift wave-like motion. For the HW system it is obtained as $\lambda_{\rm Rh} \propto \sqrt{u_{E\times B}}$, leading to a characteristic turnover time $\tau_{\rm turn} \sim \lambda_{\rm Rh} / u_{E\times B} \propto 1/ \sqrt{u_{E\times B}}$. The avalanching behavior occurs, when the eddy turnover time, $\tau_{\rm turn}$, is small compared with the instability growth time. For the purpose of formal ordering, we require $\tau_{\rm turn}\ll \gamma_{\rm L}^{-1}$, where $\gamma_{\rm L}$ is the linear growth rate. Using here that the linear growth rate is proportional with nonadiabaticity of the fluctuations, i.e., $\gamma_{\rm L} \propto \delta$, we can order, up to numerical coefficients, 
\begin{equation}
\delta\tau_{\rm turn}\propto \delta / \sqrt{u_{E\times B}} \ll 1.
\label{Rh} % Eq.~(\ref{Rh})
\end{equation} 
One sees that the turbulence driving rate, represented by the parameter $\delta$, must be small compared with the inverse Rhines time in the system. The time scale separation in Eq.~(\ref{Rh}) is favored in the regime of strong coupling or near adiabaticity on the one hand, and in the presence of large electrostatic fluctuations or fast $E\times B$ drifts, on the other hand. In the latter case the restrictive assumptions of smallness of fluctuations underlying the HW model are invalidated. Furthermore, since the instability drive of the turbulence is controlled by electron-ion collisions in the parallel electron motion via $\delta \propto \nu_{ei} / k_{||}^2$, the condition for time scale separation in Eq.~(\ref{Rh}) implies that the particle cross-field transportation time is short compared with the characteristic resistive time. Thus, the avalanches tend to dissipate their content mainly upon reaching the boundaries and merely redistribute the particles and free energy across the system otherwise. These conditions of slow driving and time scale separation combined with the stabilizing role of boundary dissipation constitute a typical set-up for dynamical systems exhibiting SOC \cite{Ash2013,SSRv}. At this point, the propagation of unstable fronts due to the processes of eddy merging and interactions acquires the typical signatures of the avalanching dynamics of the SOC type.

\section{Boundary dissipation and the role of feedback} Nonlinearly, the absorbing boundary at the plasma edge provides a feedback of the particle and energy loss processes on the dynamical state of bulk plasma; where the system of gradients pumping the drift-waves self-adjusts to generate self-organization in a marginally stable state. Thus, SOC attracts nonlinear feedback dynamics. In the meanwhile, the condition for time scale separation in Eq.~(\ref{Rh}) suggests that conservation of the transported quantity is necessary to obtain criticality. All {conservative} SOC models share this basic feature \cite{MF98,NJP,Zhang}.{\footnote {In some models of SOC, however, the assumption that the transported quantity is conserved can be relaxed, as for instance in the model introduced by Manna {\it et al}. \cite{Manna}. This model is characterized by the fraction of energy that dissipates from the system during each relaxation event. Also in some applications of SOC, such as solar flares, e.g., Refs. \cite{SSRv,Flare}, self-organization processes operate on networks, low-dimensional structures, whose topological boundary is, essentially, the network itself. In those cases, bulk dissipation is naturally included, and is consistent with critical behavior, since it acquires via the low dimensionality the precise role of boundary dissipation in conservative SOC models.} Generally, SOC can occur in every system with a negative feedback mechanism \cite{Ash2013}. Introducing an analogy with traditional thermodynamics, a thermostat is designed in such a way that the same temperature is maintained through fluctuations, so that the system is self-controlling. When, on top of this, the system possesses many degrees of freedom with next-neighbor coupling, and the driving rate is taken to be infinitesimally slow, then multi-scale behaviors will emerge, leading to fractal patterns and a $1/f$ noise of the fluctuations \cite{Bak}. %

Even so, the primary element to SOC is nonlinear feedback dynamics \cite{Kadanoff,Sor} and not the shape of the noise spectrum. It is therefore licit to claim SOC for finite-size systems with small but finite driving rates (i.e., $\delta \ll 1$) such as magnetically confined plasma systems, keeping in mind the natural limitations of associate power-law behavior. Indeed spectra resembling a $1/f$ type noise have been observed in edge plasma fluctuations and their connection with SOC has been also discussed \cite{Newman96,Tokunaga,Carr1,Pedrosa99,Rhodes,Hahm,Carr99}. 

The observational consequence of feedback lies in the fact that the average gradient controlling the instabilities will be attracted to a critical (marginally unstable) slope, where the system is essentially very sensitive to the driving \cite{Carr2006}. Nonlinearly, when a saturated turbulence state is approached, the initial fine-scale signatures of the fluctuations will be washed out by the amplification. When this occurs, the dynamics are dominated by the bulk-average nonlinearities, and not anymore by multi-scale features of fluctuations in the HW regime, permitting appreciable departures from the state of marginal stability. This behavior bears signatures, enabling to associate it with a class of bursting the so-called ``fishbone-like" instabilities \cite{Milo13,NJP} of SOC. Analysis of these strongly nonlinear regimes with feedback requires global models accounting for a self-consistent evolution of the background density profiles on an equal footing with the multi-scale fluctuation dynamics. 

\section{Stimulated vortex formation and the Galton-Watson chain process} Here we refine the avalanching model above and we employ the idea of {\it stimulated vortex formation} to assess the L\'evy index $\mu$ used to generalize the Laplacian. We think of the turbulence as starting from a uniform fluctuation background within which there is some small probability of spontaneous vortex formation. Then the birth of one vortex will perturb the unstable background nearby, such that one or more vortices will appear next to the mother vortex. In two dimensions, the energy cascading from smaller to larger scales will cause the mother and the off-spring vortices to merge, and the process will repeat itself. This process of {stimulated} vortex formation{\footnote {The idea that activity in one region can stimulate activity in another region, particularly in a nonlinear context, is in fact very general and as such must occur in many applications. Here we mention the processes of stimulated galaxy formation discussed by Schulman and Seiden \cite{Schulman}, who used this to model the hierarchical structure in the distribution of galaxies with power-law correlations.}} will generate an ever-amplifying instability front until the transferred energy is let out through the boundaries. The amplification of the vortical motions finds its energy reservoir in the nonadiabatic properties of the fluctuations, characterized by the parameter $\delta$. We assume that the original process of the vortex formation occurs near marginality, that is, near the instability threshold, and that it is self-reproducing. Then it can be modeled as a variant of the Galton-Watson chain process near extinction \cite{Clos}. This process will be self-similar in $2+1$ dimensions, with time interpreted as the preferred dimension (see Fig.~1). Note that the ``arrow of time" in the processes which we discuss looks in the direction of the energy transfer (the direction of amplification). So the time coordinate plays a very special role in the model in that it introduces in a parametric form a subordination to the hierarchy of vortices. The spatio-temporal character of the Galton-Watson process matches with the implication of SOC, where the space correlations in hierarchic geometry act as attracting the nonlinear feedback dynamics \cite{Sor}. %

\begin{figure}
\includegraphics[width=0.52\textwidth]{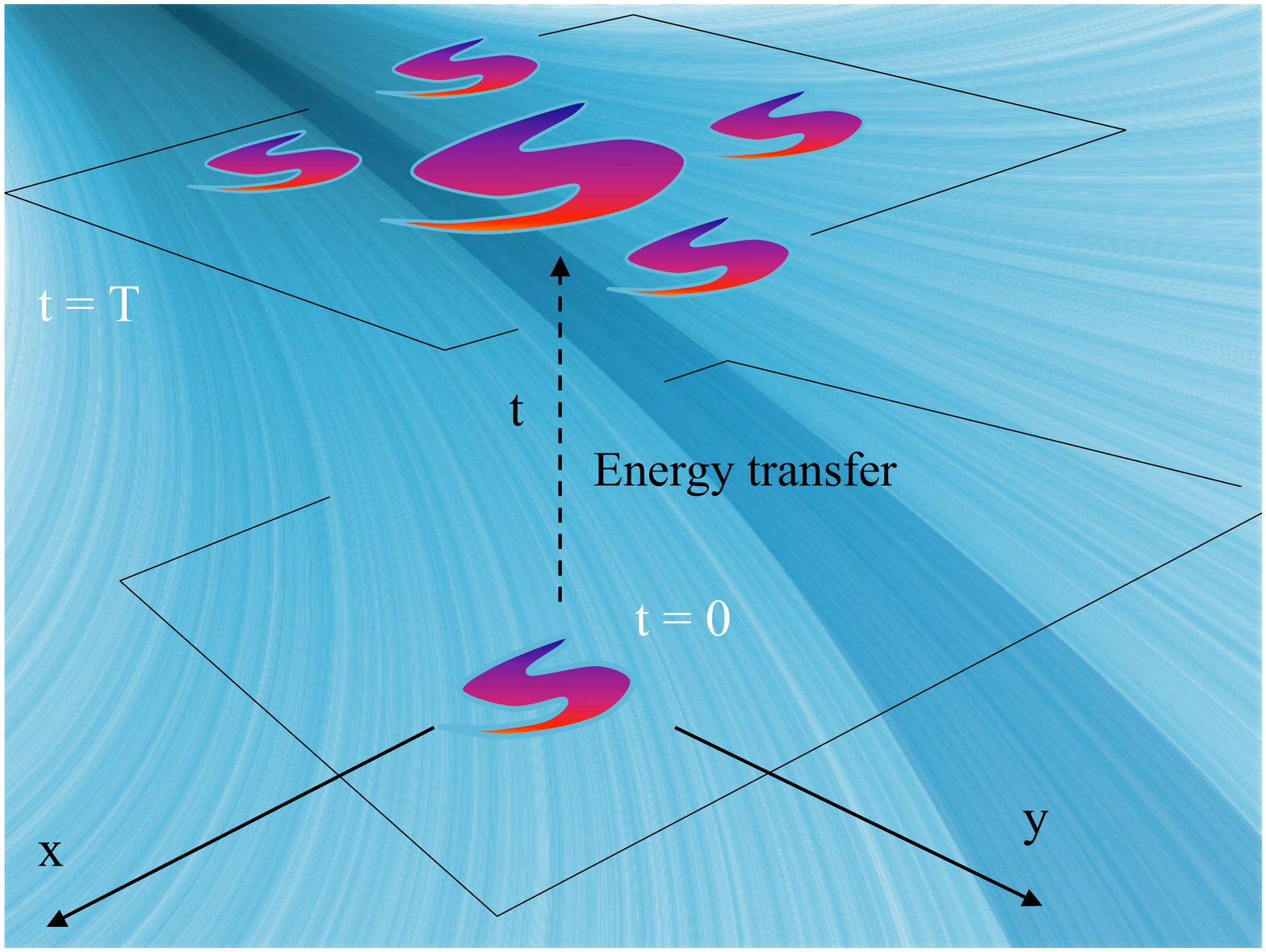}
\caption{\label{} Stimulated vortex formation as a Galton-Watson chain process in $2+1$ embedding dimensions, with time, $t$ interpreted as the preferred dimension. An artist's view, using a continuum background to accentuate the spatio-temporal character of the dynamics behind the phenomena of SOC. The ``arrow of time" looks in the direction of energy transfer.}
\end{figure}

A hierarchic process being the Galton-Watson process near extinction implies that the size distribution of its branches (i.e., eddy size distribution in the extended space) is given by a pure power-law $\hat \chi (\lambda) \sim \lambda^{-\tau + 1}$, with the exponent $\tau=5/2$ exactly. The latter exponent is obtained using binomial distribution and the Catalan numbers \cite{Clos}. This value is robust in that it does not depend on the specific version of the Galton-Watson process that is employed. We note in passing that the scaling behavior $\hat \chi (\lambda) \sim \lambda^{-3/2}$ naturally satisfies the Sparre Andersen universality \cite{SA53} and is consistent with the absorbing boundary condition in Eq.~(\ref{Sparre}). The result $\tau=5/2$ differs from the analogous value (i.e., $\tau\approx 2.33$ in three dimensions) in the original Bak, Tang and Wiesenfeld sandpile SOC model \cite{Tang}, suggesting a different universality class \cite{Milo13,Maslov}. In the above we used hat to indicate that the number density $\hat \chi (\lambda)$ is defined in $(2+1)$-dimensional space. In the limit of long times (small wave vectors), however, we can factorize the dependencies in $\hat \chi (\lambda)$ to obtain $\chi (\lambda) \sim \hat\chi (\lambda) / \lambda \propto \lambda^{-\tau}$, where $\chi (\lambda)$ is the usual eddy size distribution in the real space. Assuming that there is a characteristic flow velocity, the $E\times B$ velocity, the distribution of eddy sizes can be translated into a heavy-tailed flux distribution, $\chi (\Gamma) \sim \Gamma^{-\tau}$. This, together with the fact that the turbulent perpendicular fluxes are determined by the trapping and subsequent displacements of particles with vortical motions, leads to a power-law distribution of jump lengths $\chi (\Delta x) \sim |\Delta x|^{-\tau}$, where the exponent of the power-law is just $-\tau$. This will be consistent with a description in terms of FFPE, if  
\begin{equation}
\mu = \tau - 1,
\label{Exp-tau} % Eq.~(\ref{Exp-tau})
\end{equation} 
where Eq.~(\ref{Jump-l}) has been considered. The end result is $\mu = 3/2$. This value clearly falls in the range of validity of statistics of the L\'evy type. In view of the above we associate this value with the phenomena of SOC-turbulence coupling. In SOC theories when mean-field behavior is evaluated (the embedding dimension goes to infinity) \cite{Milo13,MF98,Tang} the $\tau$ exponent is shown to take its limiting value $\tau_{\rm MF}=3$. Then Eq.~(\ref{Exp-tau}) reveals a cross-over to the diffusive scaling, i.e., $\mu_{\rm MF}=2$, as it should. Further the FFPE in Eq.~(\ref{FFPE}) dictates the following dispersion law for the transport: $t\sim |x|^\mu$, from which the Hurst exponent $H = 1/\mu$ can be deduced. With the use of $\mu = 3/2$ one obtains $H = 2/3$ indicating superdiffusion. This result is in quantitative agreement with the results of computer simulations of cold pulse propagation in Ref. \cite{Pulse}. In this regard, we note that values of the Hurst exponent numerically close to $H\approx 0.7$ are found phenomenologically in a remarkably broad range of applications \cite{Feder}, spanning from fluctuations of the river Nile to the dynamics of edge plasma turbulence \cite{Carr1,Pedrosa99}.    

\section{From log-normal behavior to Pareto-L\'evy tail} That the processes of stimulated vortex formation along with the amplification processes in the presence of inverse energy cascade will lead to a heavy-tailed distribution of jump lengths and fluxes finds further support in the general properties of log-normal behavior. It is noticed, following Ref. \cite{PLA}, that the flux-surface averaged flux, $\Gamma = \int d\sigma \left[\tilde n u_{E\times B}\right]/ \int d\sigma$, is to lowest order positive definite, and that it has multiplicative character in that it involves a product of two random quantities, the fluctuating particle density, $\tilde n$, and the fluctuating $E\times B$ velocity. Then the ``central limit theorem" of the theory of the probability will imply that it is the logarithm of the averaged flux, which is normally distributed.{\footnote {It is implicit in this kind of reasoning that the fluctuations are small in a sense, so that cross-correlation between the density and the velocity fluctuations can be neglected. Then the smallness of the fluctuations makes it possible to rely on the HW picture of drift-wave turbulence leading to the probability density functions with exponentially decaying tails.}} Indeed it is found in direct numerical simulations \cite{PLA} of the HW model that the probability distribution functions of the flux-surface averaged transport agree well with a log-normal distribution (see Fig.~2). These results also generalize to electromagnetic drift-Alfv\`en turbulence with magnetic field curvature effects, as well as to magnetohydrodynamic edge plasma turbulence \cite{Garcia,Garcia+}. 

\begin{figure}
\includegraphics[width=0.52\textwidth]{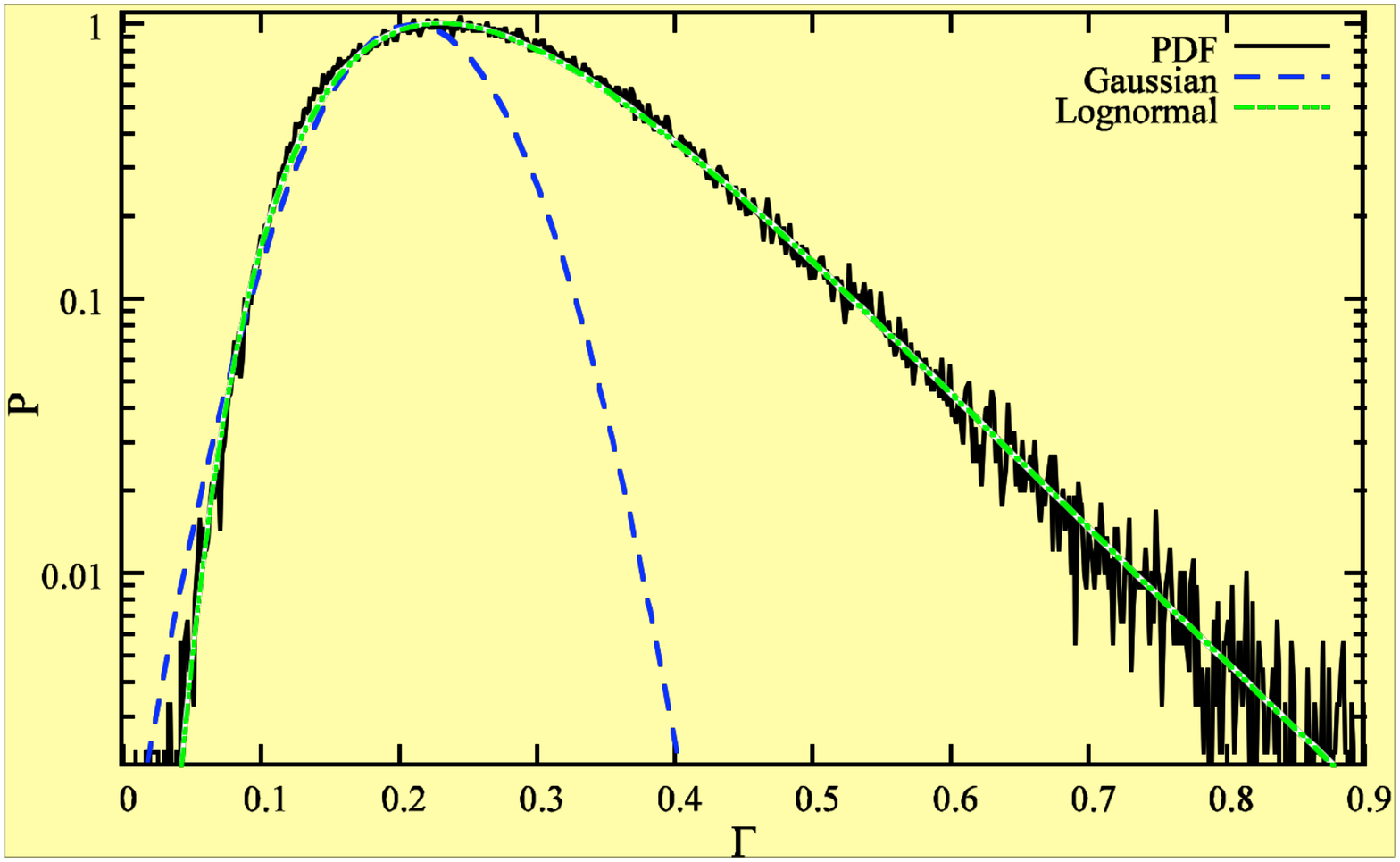}
\caption{\label{} Probability distribution functions (PDF's) of the flux-surface averaged plasma flux compared with a log-normal distribution and a Gaussian distribution from a numerical simulation of the HW system \cite{PLA}. Non-Gaussianity, which is clear from the deviation between the actual PDF and the parabolic shaped dashed line, is imposed by the multiplicative character of the flux, involving a product of two random variables, the fluctuating particle density, and the fluctuating velocity. In the absence of amplification (small density and velocity fluctuations, characterizing the HW regime) a good agreement with log-normal distribution is found, consistently with the implication of the central limit theorem of the theory of the probability.  
%Adapted from Ref. \cite{PLA}. 
}
\end{figure}

As Montroll and Shlesinger realized \cite{Montroll}, any initially log-normal distribution will change to a distribution without well-defined moments when/if its mean value is unlimitedly amplified by some process. With the aid of a recursion relation it was argued that the new distribution that allows for these amplifications has a nonanalytic part leading to the Pareto-L\'evy tail in the sense of a $\mu$-stable L\'evy motion and generalized central limit theorem \cite{Gnedenko}. Interestingly, the above authors motivated their study with explanation of $1/f$ noise and with examples involving hierarchical random processes with subordination. They suggested that, when subordination occurs in many orders, the distribution function of successes in the primary order is log-normal. Using here that the inverse energy cascade, characteristic of two-dimensional fluid turbulence, acts as to introduce a ``subordination" into the hierarchy of turbulent eddies, parametrized by their wave number, and that the avalanches themselves emerge from same turbulence vortex motions provided just that there is a time scale separation, $\delta / \sqrt{u_{E\times B}} \ll 1$, one readily concludes that the amplification processes taking place will naturally generate the wanted Pareto-L\'evy inverse-power tail, $\chi (\Gamma) \sim \Gamma^{-1-\mu}$, where the exponent $\mu$ characterizes the hierarchy of amplification. We associate the $\mu$ value with the Galton-Watson chain process discussed above. 

Let us now take stock at this point and summarize our results so far: Space-fractional derivatives enter the FFPE in Eq.~(\ref{FFPE}) very nontrivially in that neither a classical SOC nor classical turbulence models can readily accommodate them as such. Then an amplification process is needed to validate nonlocal models of anomalous transport. The mechanism of amplification has referred to a complexity coupling between SOC and drift-wave turbulence phenomena in the presence of nonlinear feedbacks at the plasma edge. We have seen in the above that the regimes with strong fluctuations in hierarchical systems are characterized by a statistics of the L\'evy type and nonanalytic distribution functions with algebraically decaying heavy tails. We consider these tails as representing the L\'evy motions in the flow. In Langevin equations~(\ref{Lvin}) we charge the corresponding derivative process, the white L\'evy noise $F_L (t)$, to generate via the transition probability in Eq.~(\ref{GCLT}) a nonlocal FFPE.

The operation of a negative feedback mechanism, other than providing a route to SOC phenomena, guarantees the necessary stiffness to profiles near a marginally stable state. It will also explain some asymmetry \cite{Pulse,Pulse+} between the propagation of perturbations due to heat modulation and cold pulses. For $x > x_s$, where $x_s$ denotes the location of ion cyclotron resonance heating power deposition, heat waves and pulses propagate fast. However, for $x < x_s$, the heat wave slows down and is damped, but the cold pulses still travel fast. The explanation lies in the fact that the application of a cold pulse to the plasma edge steepens the gradient of the average profile, thus turning it into the unstable (supercritical) domain. Then the system responds by plasma instabilities and the phenomena of turbulent amplification of the fluxes. So, the transport is nonlocal, and the transport problem for the cold pulse is essentially a L\'evy flight problem. By contrast, the application of heat power modulation introduces some sort of knee to the profile. Indeed the profile becomes steeper (and thus unstable) in the range $x > x_s$ where behavior is supercritical involving nonlocality, and flatter in the range $x < x_s$ which is subcritical and which damps the wave. 

We should stress that the explanation of cold pulse behavior involves a critical threshold condition and does not pertain to the usual Fickian transport paradigm. These properties are reflected by the FFPE in Eq.~(\ref{FFPE+}), where the intensity $D_\mu$ is only nonzero at or above a critical gradient generating the noise process of the L\'evy type. 

\section{Spectral energy density} Based on the above analysis, we might suggest that the mechanism of turbulent amplification of the SOC avalanches is responsible for the occurrence of large-amplitude transport events in plasma confinement, associable with large intermittent bursts of transport \cite{Ippolito,Xu10}. One question of practical importance is concerned with the precursors of this behavior. Because of amplification, we expect a steeper drop-off in the energy spectrum of the coupled SOC-turbulence system as compared to the inertial range of the fluid (drift-wave) turbulence. Using here that the divergence of the $E\times B$ drift is zero, which keeps the flow incompressible, one finds after many merging events that the variance ${\rm Var}\,u_{E\times B}\, (\lambda) \equiv \langle|u_{E\times B}(x+\lambda) - u_{E\times B}(x)|^2\rangle$ will be proportional with the square of the vortex size, leading to ${\rm Var}\,u_{E\times B}\, (\lambda) \propto \lambda^2$. When translated into wave vectors, $\lambda\sim 1/k$, this becomes ${\rm Var}\,u_{E\times B}\, (k) \propto k^{-2}$. Employing for the spectral energy density, $S(k)$,  
\begin{equation}
{\rm Var}\,u_{E\times B}\, (\lambda) = 2\int_{k_{\min}}^{k_{\max}} {{S}} (k^{\prime}) \left[1-\frac{\sin k^{\prime} \lambda}{k^{\prime}\lambda}\right] dk^{\prime}
\label{Disper} % Eq.~(\ref{Disper})
\end{equation} 
one readily obtains $S(k) \sim k^{-3}$. Indeed the spectrum of measured fluctuations in tokamak plasmas involves a $k^{-3}$ power-law subrange already discussed in Ref. \cite{Misguish}. By contrast, assuming that the energy cascade rate does not depend on $k$, one arrives at the scaling ${\rm Var}\,u_{E\times B}\, (k) \propto k^{-2/3}$, from which the familiar Kolmogorov spectrum $S(k) \sim k^{-5/3}$ is inferred. In this connection, we should stress that the avalanching transport is triggered by the explicit radial dependence in the profiles, and, when account is taken for the inverse cascade, by boundary feedbacks, so that the assumptions of constant energy transfer and of infiniteness of the system, resulting in the fluid-like $-5/3$ behavior, do not really apply here. 

\section{Conclusions} We have proposed a combined model of nonlocal transport which brings the notion of inverse energy cascade, typical of drift-wave- and two-dimensional fluid turbulence, in contact with the ideas of avalanching dynamics, characteristic of SOC. The new model, which we discuss, was motivated by the study of the cold pulse propagation in magnetically confined toroidal plasma \cite{Pulse}, although the emphasis of this paper is entirely on the fundamental aspects of nonlocal behavior. We have seen in the above that a L\'evy fractional Fokker-Planck equation is neither consistent with the classical drift-wave- (HW style) and two-dimensional fluid turbulence approach, nor with a classical SOC approach built on the assumptions of locality and next-neighbor interactions, and that an additional amplification process is needed to bridge the gaps between the theoretical concepts of fluid turbulence, SOC and nonlocal FFPE. 

We suggest that the amplification occurs via a complexity coupling between the phenomena of drift-wave (two-dimensional fluid) turbulence and SOC in the presence of an absorbing boundary at the plasma edge. It requires a strong nonlinearity in that the Rhines time $\tau_{\rm Rh} \propto 1/ \sqrt{u_{E\times B}}$ must be small compared with the instability growth time. Whereas the boundaries can, in the thermodynamic limit, be assumed at infinity, they are an essential key element to the model as they introduce a feedback dynamics generating SOC. Then there is a theoretical possibility that the turbulence fuels the avalanching dynamics due to SOC through inverse cascade of the energy, giving rise to transport events of anomalously large size beyond the range of predictability of the ``conventional" SOC. The energy reservoir for this behavior is only limited to the size of the confinement system. The phenomenon has serious implications for operational stability of big confinement devices such as the fusion power plants, where it may trigger off transport events of potentially a catastrophic character. These theory predictions being rather alarming might become the ``inconvenient truth" of the fusion research. Other than fusion, the phenomena of SOC-turbulence coupling might be proposed for geophysical flows (using as appropriate the notion of Rhines length), where they shall be responsible for outstanding perturbations beyond the expected weather and climate patterns. Statistically, the big events will manifest themselves in the form of Pareto-L\'evy tails on top of the typical log-normal probability distribution function of the flux-surface averaged transport. This suggestion finds its justification in the general properties of log-normal behavior in hierarchical systems with subordination \cite{Montroll}. 

It was argued that SOC was not really an alternative to the notion of turbulence in that it operates as a dynamically induced nonlinear twist in basically a turbulent medium. This nonlinearity is implicitly present in the value of the fractional exponent $\mu$ used to generalize the Laplacian. It leads to a nontrivial situation, in which the transport equation, the FFPE, is formally linear, with feedback nonlinearities absorbed by the fractional order of space differintegration. The exponent of fractional differintegration $\mu$ determines the anomalous scaling between the lifetime of the avalanches and their size, i.e., $t\sim |x|^\mu$. This scaling relation is often found phenomenologically in fluctuations of the low confinement mode plasma, where it is associated with a pure SOC \cite{Politzer}. Indeed the same scaling relation is obtained from major SOC models where it derives from the assumptions of locality and next-neighbor interactions. In those settings the exponent $\mu$ which is model dependent determines the universality class of SOC \cite{Milo13,Maslov}. 

There is no indication that our mixed SOC-turbulence model falls within the known universality classes. Employing the notion of the Galton-Watson chain process, and the idea of stimulated vortex formation for electrostatic drift-wave turbulence, we find $\mu = 3/2$ exactly. The model observes the Sparre Andersen universality \cite{SA53} in the presence of an absorbing boundary at the plasma edge. In this connection, we note that the observation of a power-law, $\chi (\Gamma) \sim \Gamma^{-1-\mu}$, does not really imply SOC. It does imply a SOC-turbulence coupling instead. 

All in all, we are led to conclude that the phenomena of drift-wave turbulence and SOC are {\it not} really separable in tokamaks in the regime of strong nonlinearity and that there is some sort of SOC-turbulence {\it duality} coming along with the time scale separation condition in Eq.~(\ref{Rh}). Identifying the ``unique" fingerprint of SOC becomes, therefore, a problem of academic interest.  

\section{Acknowledgements}
%{\it Acknowledgement:} 
We thank Volker Naulin for discussion and comments to the manuscript. %This work was supported by the Euratom Communities under the contract of Association between Euratom/ENEA and Euratom/DTU. 
This study was made possible by the EURATOM mobility scheme. Partial support was received from the ISSI (International Space Science Institute) project ``SOC and Turbulence" (Bern, Switzerland).



%% References


\bibliographystyle{elsarticle-num}
\bibliography{<your-bib-database>}






% \end{thebibliography}

\end{document}